\documentclass[11pt]{article}
\usepackage{modappb,epsfig,axodraw}

\def\ie{i.e.\ }
\def\eg{e.g.\ }
\def\cf{cf.\ }

\newcommand{\real}{{\cal\mbox{Re\,}}}

\def\beq{\begin{equation}}
\def\eeq{\end{equation}}
\def\beqar{\begin{eqnarray}}
\def\eeqar{\end{eqnarray}}
\def\barr#1{\begin{array}{#1}}
\def\earr{\end{array}}
\def\bfi{\begin{figure}}
\def\efi{\end{figure}}
\def\btab{\begin{table}}
\def\etab{\end{table}}
\def\bce{\begin{center}}
\def\ece{\end{center}}

\def\text{\textstyle}

\def\refeq#1{\mbox{(\ref{#1})}}
\def\reffi#1{\mbox{Fig.~\ref{#1}}}

\def\citere#1{\mbox{Ref.~\cite{#1}}}
\def\citeres#1{\mbox{Refs.~\cite{#1}}}

\def\al{\alpha}
\def\be{\beta}
\def\ga{\gamma}
\def\de{\delta}
\def\om{\omega}

\def\si{\sigma}
\def\Ga{\Gamma}
\def\De{\Delta}

\newcommand{\lsim}
{\;\raisebox{-.3em}{$\stackrel{\displaystyle <}{\sim}$}\;}

\newcommand{\GeV}{\unskip\,\mathrm{GeV}}
\newcommand{\MeV}{\unskip\,\mathrm{MeV}}

\def\mathswitchr#1{\relax\ifmmode{\mathrm{#1}}\else$\mathrm{#1}$\fi}
\def\mathswitch#1{\relax\ifmmode#1\else$#1$\fi}

\newcommand{\PV}{\mathswitch V}
\newcommand{\PW}{\mathswitchr W}
\newcommand{\PZ}{\mathswitchr Z}

\newcommand{\PH}{\mathswitchr H}
\newcommand{\Pe}{\mathswitchr e}
\newcommand{\Pne}{\mathswitch \nu_{\Pe}}
\newcommand{\Pane}{\mathswitch \bar\nu_{\Pe}}
\newcommand{\Pnmu}{\mathswitch \nu_\mu}
\newcommand{\Panmu}{\mathswitch \bar\nu_{\mu}}
\newcommand{\Pnta}{\mathswitch \nu_\tau}
\newcommand{\Panta}{\mathswitch \bar\nu_{\tau}}

\newcommand{\Pad}{\mathswitch \bar{\mathrm{d}}}

\newcommand{\Pu}{\mathswitchr u}

\newcommand{\Pep}{\mathswitchr {e^+}}
\newcommand{\Pem}{\mathswitchr {e^-}}

\newcommand{\MV}{\mathswitch {M_\PV}}
\newcommand{\MW}{\mathswitch {M_\PW}}

\newcommand{\MZ}{\mathswitch {M_\PZ}}
\newcommand{\MH}{\mathswitch {M_\PH}}
\newcommand{\Me}{\mathswitch {m_\Pe}}
\newcommand{\Mmy}{\mathswitch {m_\mu}}

\newcommand{\GW}{\mathswitch {\Gamma_\PW}}
\newcommand{\GV}{\mathswitch {\Gamma_\PV}}

\newcommand{\kV}{\mathswitch {p_\PV}}

\newcommand{\scrs}{\scriptscriptstyle}

\newcommand{\cw}{\mathswitch {c_{\scrs\PW}}}

\newcommand{\ri}{{\mathrm{i}}}

\newcommand{\rd}{{\mathrm{d}}}
\newcommand{\SU}{{\mathrm{SU}}}

\newcommand{\M}{{\cal {M}}}

\newcommand{\born}{{\mathrm{Born}}}
\newcommand{\nf}{{\mathrm{nf}}}

\begin{document}
\title{Radiative corrections to off-shell gauge-boson pair 
       production\thanks{Talks presented at the Zeuthen Workshop on
         Elementary Particle Theory: ``Loops and Legs in 
       Gauge Theories'', Rheinsberg, Germany, April 19\,--\,24, 1998.}
}
\author{Wim Beenakker
        \address{Instituut--Lorentz, University of Leiden, NL--2300 RA Leiden,
                 The Netherlands}
        \and Ansgar Denner
        \address{Paul Scherrer Institut, CH--5232 Villigen PSI, Switzerland}
}
\maketitle

\begin{abstract}
We give an overview of the problems and developments associated with the
calculation of radiative corrections to off-shell gauge-boson pair
production in $\Pep\Pem$ collisions.
\end{abstract}

\section{Introduction}

In recent years we have grown increasingly accustomed to the great success of 
the Standard Model (SM) of electroweak interactions. However, up to now only a 
relatively restricted sector of the SM has been checked. The Yang--Mills
character of the gauge-boson self-interactions and the Higgs mechanism of mass
generation still await conclusive confirmation. This leaves ample of room 
for interesting physics at upcoming collider experiments, either within or
outside the SM. If any physics beyond the SM exists, it will reveal itself in 
the production of new particles (direct signals) or in deviations in the
interactions between the SM particles (indirect signals). 

Reactions that involve the production and subsequent decay of pairs of 
unstable gauge bosons ($\Pep\Pem\to V_1 V_2\to 4f+n\ga$) constitute powerful 
search tools for such indirect signals. In this context, the limelight is 
presently on the production of pairs of W bosons at LEP2, the second stage of 
the LEP program. With energies above the nominal W-pair-production threshold, 
LEP2 offers a twofold possibility of testing the SM. First of all, it allows a
precise direct measurement of the W-boson mass (\MW) with an envisaged 
precision of $\De\MW =$ 40\,--\,50\,MeV~\cite{LEP2MW}. By combining this 
precise measurement of \MW\ with the high-precision data on $\al$, $G_{\mu}$, 
and \MZ, the mass of the top quark can be calculated within the SM as a 
function of the Higgs-boson mass \MH\ and the strong coupling 
$\al_s$~\cite{LEP2WW}. The so-obtained top-quark mass can then be confronted 
with the direct bounds from the Tevatron and the indirect ones from the 
precision measurements at LEP1/SLC. In this way improved limits on \MH\ can be 
obtained. On the other hand, LEP2 can provide information on the 
structure of the triple gauge-boson couplings (TGC)~\cite{LEP2TGC}. These 
couplings appear at tree level in LEP2 processes like $\Pep\Pem \to 4f$ or 
$\Pep\Pem \to \Pne\Pane\ga$, in contrast to LEP1 observables where they only 
entered through loop corrections. At LEP2 one hopes to determine the TGC with 
a relative precision of $\,\sim 10\%$ with respect to the SM couplings.

At a high-energy linear collider (LC), with its envisaged energy in the range 
500\,--\,2000\,GeV, one can go one step further. Apart from TGC studies well
below the per-cent level, the high LC energies also open the possibility 
of studying quartic gauge-boson couplings in reactions like 
$\Pep\Pem \to 6f$ or $\ga\ga \to 4f$, thereby entering the realm of the 
symmetry-breaking mechanism~\cite{NLCTGC}. On top of that one can look for 
signs of a strongly-interacting symmetry-breaking sector (\ie resonances or 
phase shifts), by studying longitudinal gauge-boson interactions in 
$\Pep\Pem \to 4f,\,6f$ and $\ga\ga \to 4f$. 

In order to successfully achieve the above-mentioned physics goals, a very 
accurate knowledge of the SM predictions for the various observables is 
mandatory. This requires a proper understanding of radiative corrections as 
well as a proper treatment of finite-width effects. In the next sections we 
address these two topics in detail.

\section{Gauge-invariant treatment of unstable gauge bosons}

\subsection{Lowest order}

The above-described physics issues all involve an investigation of processes
with photons and/or fermions in the initial and final state and
unstable gauge bosons as intermediate particles. If complete sets 
of graphs contributing to a given process are taken into account, the 
associated matrix elements are in principle gauge-invariant, \ie they are
independent of gauge fixing and respect Ward identities. 
However, the gauge bosons that appear as intermediate particles can give rise 
to poles $1/(\kV^2-\MV^2)$ in physical observables if they are treated as 
stable particles. This can be cured by introducing the finite decay width.
In field theory, such widths arise naturally from the imaginary 
parts of higher-order diagrams describing the gauge-boson self-energies, 
resummed to all orders. However, in doing a Dyson summation of self-energy 
graphs, we are singling out only a very limited subset of all possible
higher-order diagrams. It is therefore not surprising that one often ends up 
with a result that violates Ward identities and/or retains some gauge
dependence resulting from incomplete higher-order contributions. 

Until a few years ago two approaches were popular in the construction
of lowest-order LEP2/LC Monte Carlo generators.  The first one, the
so-called ``fixed-width scheme'', involves the systematic replacement
$1/(\kV^2-\MV^2) \to 1/(\kV^2-\MV^2+\ri\MV\GV)$, where \GV\ denotes
the physical width of the gauge boson with mass \MV\ and momentum \kV.
Since in perturbation theory the propagator for space-like momenta
does not develop an imaginary part, the introduction of a finite width
also for $\kV^2<0$ has no physical motivation and in fact violates
unitarity, \ie the cutting equations. This can be cured by using a
running width $\ri\MV\GV(\kV^2)$ instead of the constant one
$\ri\MV\GV$ (``running-width scheme'').

As in general the resonant diagrams are not gauge-invariant by themselves, 
the introduction of a constant or running width destroys gauge invariance.  
At this point the question arises whether the gauge-breaking terms are 
numerically relevant or not. After all, the gauge breaking is caused by the 
finite decay width and is, as such, in principle suppressed by powers of 
$\GV/\MV$. For LEP1 observables we know that gauge breaking can be negligible 
for all practical purposes. However, the presence of small scales can amplify 
the gauge-breaking terms. This is for instance the case for almost collinear 
space-like photons~\cite{BHF1} or longitudinal gauge bosons ($\PV_L$) at high 
energies~\cite{BHF2}, involving scales of ${\cal O}(p_{_B}^2/E_{_B}^2)$ for
$B=\gamma,\PV_L$. The former plays an important 
role in TGC studies in the reaction $\Pep\Pem \to \Pem\Pane\Pu\Pad$, where the
electron may emit a virtual photon with $p_{\ga}^2$ as small as $\Me^2$. 
The latter determines the high-energy behaviour of the generic reaction
$\Pep\Pem \to 4f$. In these situations the external current coupled to the 
photon or to the longitudinal gauge boson becomes approximately proportional
to $p_{_B}$. Sensible theoretical predictions, with a proper dependence on 
$p_{\ga}^2$ and a proper high-energy behaviour, are only possible if the 
amplitudes with external currents replaced by the corresponding gauge-boson 
momenta fulfill appropriate Ward identities. 

In order to substantiate these statements, a truly gauge-invariant scheme is
needed. It should be stressed, however, that any such scheme is arbitrary to a 
greater or lesser extent: since the Dyson summation must necessarily be taken 
to all orders of perturbation theory, and we are not able to compute the 
complete set of {\it all} Feynman diagrams to {\it all} orders, the various 
schemes differ even if they lead to formally gauge-invariant results. Bearing 
this in mind, we need besides gauge invariance some physical motivation for 
choosing a particular scheme. In this context two options can be mentioned.
The first option is the so-called ``pole scheme''~\cite{pole-scheme}.
In this scheme one decomposes the complete amplitude by expanding around the 
poles. As the physically observable residues of the poles are gauge-invariant, 
gauge invariance is not broken if the finite width is taken into account in 
the pole terms $\propto 1/(\kV^2-\MV^2)$. It should be noted that there is 
no unique definition of these residues. Their calculation involves a mapping 
of off-shell matrix elements with off-shell kinematics on on-resonance matrix 
elements with restricted kinematics. The restricted kinematics, however, is 
not unambiguously fixed. After all, it contains more than just the invariant 
masses of the unstable particles and one has to specify the variables that have
to be kept fixed when determining the residues. The resulting implementation 
dependence manifests itself in differences of subleading nature, 
\eg ${\cal O}(\GV/\MV)$ suppressed deviations in the leading pole-scheme
residue. In particular near phase-space boundaries, like thresholds,
the implementation differences can take on noticeable proportions.

We roughly sketch the pole-scheme method for a single unstable particle.
The Dyson resummed lowest-order matrix element is given by
\beqar
  \label{pole-scheme}
  \M^\infty &=& \frac{W(\kV^2,\om)}{\kV^2-\MV^2}\,\sum_{n=0}^{\infty}\,
                \Biggl( \frac{\Sigma_\PV(\kV^2)}{\kV^2-\MV^2} \Biggr)^n 
                =\ \frac{W(\kV^2,\om)}{\kV^2-\MV^2-\Sigma_{\PV}(\kV^2)} \\
            &=& \frac{W(M^2,\om)}{\kV^2-M^2}\,\frac{1}{Z(M^2)}
                + \Biggl[ \frac{W(\kV^2,\om)}{\kV^2-\MV^2-\Sigma_{\PV}(\kV^2)}
                      - \frac{W(M^2,\om)}{\kV^2-M^2}\,\frac{1}{Z(M^2)} \Biggr],
                \nonumber
\eeqar
where $\Sigma_\PV(\kV^2)$ is the self-energy of the unstable particle, 
$M^2$ is the pole in the complex $\kV^2$ plane, and $Z(M^2)$ is the 
wave-function factor: $\,M^2-\MV^2-\Sigma_\PV(M^2) = 0\,$ and 
$\,Z(M^2) = 1-\Sigma'_{\PV}(M^2)\,$. The first term in the last expression of
\refeq{pole-scheme} represents the single-pole residue, which is closely 
related to on-shell production and decay of the unstable particle. The argument
$\om$ denotes the dependence on the other variables, \ie the implementation
dependence. The second term has no pole and can be expanded in powers of 
$\kV^2-M^2$.

The second option is based on the philosophy of trying to determine and 
include the minimal set of Feynman diagrams that is necessary for compensating 
the gauge violation caused by the self-energy graphs. This is obviously the 
theoretically most satisfying solution, but it may cause an increase in the 
complexity of the matrix elements and a consequent slowing down of the 
numerical calculations. For the gauge bosons we are guided by the observation 
that the lowest-order decay widths are exclusively given by the imaginary 
parts of the fermion loops in the one-loop self-energies. It is therefore 
natural to perform a Dyson summation of these fermionic one-loop self-energies 
and to include the other possible one-particle-irreducible fermionic one-loop 
corrections (``fermion-loop scheme'')~\cite{BHF1,BHF2}. For the LEP2 process 
$\Pep\Pem \to 4f$ this amounts to adding the fermionic corrections to
the triple gauge-boson vertex. The complete set of fermionic
contributions forms a gauge-independent subset and obeys all Ward
identities exactly, even with resummed propagators~\cite{BHF2}.

Having two gauge-invariant calculational schemes to compare with, we can now
have a closer look at the ever-popular fixed- and running-width
schemes. The running-width scheme violates electromagnetic and $\SU(2)$ gauge
invariance already in $\Pep\Pem \to 4f$ and is found to produce
completely unreliable results~\cite{BHF1,BHF2}. Within the fixed-width scheme
electromagnetic gauge invariance is preserved in $\Pep\Pem \to 4f$ 
for massless external particles, eliminating problems with almost collinear 
space-like photons. This is indeed confirmed by the numerical comparison in
\citeres{BHF1,BHF2}. In more general reactions, like $\Pep\Pem \to 4f\ga$,
electromagnetic gauge invariance is broken. Owing to the presence of 
non-transverse W-boson contributions, electromagnetic gauge invariance in
$\Pep\Pem \to 4f\ga$ can be achieved only if the process $\Pep\Pem \to 4f$ is 
$\SU(2)$ gauge-invariant. 
At high energies ($E_{\PV}\gg \MV$) the fixed-width scheme in general violates 
$\SU(2)$ gauge invariance by terms of the order $\MV\GV/E_{\PV}^2$~\cite{BHF2}.
In matrix elements for physical processes the gauge-invariance-violating terms 
are enhanced by a factor $E_{\PV}/\MV$ for each effectively longitudinal gauge 
boson. The process $\Pep\Pem \to 4f$ involves at most two longitudinal gauge 
bosons, such that the $\SU(2)$ gauge violation is suppressed by $\GV/\MV$ at 
the matrix-element level. Therefore, the high-energy behaviour of the 
cross-section in the fixed-width scheme is consistent with unitarity, which is
confirmed by the numerical comparison in \citere{BHF2}. 
However, our argument implies a bad high-energy behaviour for processes with 
more intermediate (longitudinal) gauge bosons, like longitudinal gauge-boson 
scattering in $\Pep\Pem \to 6f$.%
\footnote{A fixed-width scheme that preserves electromagnetic and $\SU(2)$ 
          gauge invariance is only possible if one uses {\it one} complex
          gauge-boson mass, everywhere. Consequently, the complex W and Z 
          masses have to be related by 
          $\MW^2 -\ri\MW\GW = \cw^2(\MZ^2 -\ri\MZ\Ga_{\PZ})$, leading to a 
          relation between the decay widths that is not supported by 
          experiment.}
   
\subsection{Radiative corrections}

The implementation of radiative corrections adds an additional level of 
complexity by the sheer number of contributions ($10^3$\,--\,$10^4$) that have
to be evaluated. 

By employing the fermion-loop scheme all one-particle-irreducible fermi\-onic 
one-loop corrections can be embedded in the tree-level matrix elements. This 
results in running couplings, propagator functions, vertex functions, etc. 
However, there is still the question about the bosonic corrections. A large 
part of these bosonic corrections, as \eg the leading QED corrections, 
factorize and can be treated by means of a convolution, using the 
fermion-loop-improved cross-sections in the integration kernels. This allows 
the inclusion of higher-order QED corrections and soft-photon 
exponentiation. In this way various important effects can be covered, 
as \eg the large negative soft-photon corrections near the nominal W-pair 
threshold, the distortion of angular distributions as a result of hard-photon 
boost effects, and the average energy loss due to radiated 
photons~\cite{LEP2WW,WWreview}. Nevertheless, the remaining bosonic corrections
can be large, especially at high energies~\cite{LEP2WW,WWreview,WWapp}.

In order to include these corrections one might attempt to extend the 
fermion-loop scheme. In the context of the background-field method~\cite{BFM1} 
a Dyson summation of one-loop bosonic self-energies can be performed without 
violating the Ward identities~\cite{BFM2}. However, the resulting matrix 
elements depend on the quantum gauge parameter at the loop level that is not 
completely taken into account. As mentioned before, the perturbation series 
has to be truncated; in that sense the dependence on the quantum gauge 
parameter could be viewed as a parametrization of the associated ambiguity. 

As an additional complication for such Dyson-summation techniques, we mention 
that a consistent calculation of the radiative corrections involves also the 
one-loop corrections to the decay widths. Since this requires the imaginary 
part of the two-loop self-energies, also other (imaginary parts of) two-loop 
corrections are needed to restore gauge invariance. An efficient way of 
overcoming this complication is still under investigation.

As a more appealing and economic strategy we discuss in the next section how 
the radiative corrections can be calculated in an approximated pole-scheme
expansion.

\section{Radiative corrections in the double-pole approximation}

The presently most favoured framework for evaluating the radiative 
corrections to resonance-pair-production processes, like W- and Z-pair 
production, involves the so-called double-pole approximation (DPA). This 
approximation restricts the complete pole-scheme expansion to the term with the
highest degree of resonance. In the case of W/Z-pair production only the 
double-pole residues are hence considered. The intrinsic error associated with
this procedure is $\alpha\GV/(\pi \MV) \lsim 0.1\%$, except far off resonance
and close to phase-space boundaries where also the implementation dependence 
of the double-pole residues can lead to enhancement factors. For this reason
the DPA is only valid a few \GV\ above the nominal (on-shell) gauge-boson-pair
threshold. 

In the DPA one can identify two types of contributions. One type comprises all 
diagrams that are strictly reducible at both unstable gauge-boson lines 
(see \reffi{double-pole}). These corrections are therefore called factorizable 
and can be attributed unambiguously either to the production of the 
gauge-boson pair or to one of the subsequent decays.
\bfi
  \bce
  \unitlength .7pt\small\SetScale{0.7}
  \begin{picture}(200,130)(0,-10)
    \ArrowLine(43,58)(25,40)        \Text(8,40)[lc]{\Pep}
    \ArrowLine(25,90)(43,72)        \Text(8,92)[lc]{\Pem}
    \Photon(50,65)(150,95){1}{16}   \Text(100,105)[]{\PW}  
    \Photon(50,65)(150,35){1}{16}   \Text(100,25)[]{\PW}
    \ArrowLine(155,98)(175,110)     \Text(190,110)[rc]{$f_1$}
    \ArrowLine(175,80)(155,92)      \Text(190,80)[rc]{$\bar{f}_1'$}
    \ArrowLine(175,50)(155,38)      \Text(190,50)[rc]{$\bar{f}_2'$}
    \ArrowLine(155,32)(175,20)      \Text(190,20)[rc]{$f_2$}
    \DashLine(75,115)(75,15){5}     \Text(40,5)[]{production}
    \DashLine(125,115)(125,15){5}   \Text(150,5)[]{decays}
    \GCirc(50,65){10}{1}
    \GCirc(100,80){10}{0.8}
    \GCirc(100,50){10}{0.8}
    \GCirc(150,95){10}{1}
    \GCirc(150,35){10}{1}
  \end{picture}
  \ece
  \caption[]{The generic structure of the factorizable
             W-pair contributions. The shaded circles indicate the 
             Breit--Wigner resonances.}
  \label{double-pole}
\efi%
The second type consists of all diagrams in which the production and/or decay 
subprocesses are not independent (see \reffi{fi:nf}). We refer to these 
effects as non-factorizable corrections (NFC). 
\bfi[b]
  \bce
  \unitlength .7pt\small\SetScale{0.7}
  \begin{picture}(120,100)(0,0)
  \ArrowLine(30,50)( 5, 95)
  \ArrowLine( 5, 5)(30, 50)
  \Photon(30,50)(90,80){2}{6}
  \Photon(30,50)(90,20){2}{6}
  \GCirc(30,50){10}{0}
  \Vertex(90,80){1.2}
  \Vertex(90,20){1.2}
  \ArrowLine(90,80)(120, 95)
  \ArrowLine(120,65)(105,72.5)
  \ArrowLine(105,72.5)(90,80)
  \Vertex(105,72.5){1.2}
  \ArrowLine(120, 5)( 90,20)
  \ArrowLine( 90,20)(105,27.5)
  \ArrowLine(105,27.5)(120,35)
  \Vertex(105,27.5){1.2}
  \Photon(105,27.5)(105,72.5){2}{4.5}
  \put(92,47){$\ga$}
  \put(55,73){\PW}
  \put(55,16){\PW}
  \end{picture}
  \quad
  \begin{picture}(120,100)(0,0)
  \ArrowLine(30,50)( 5, 95)
  \ArrowLine( 5, 5)(30, 50)
  \Photon(30,50)(90,80){2}{6}
  \Photon(30,50)(90,20){2}{6}
  \Vertex(60,35){1.2}
  \GCirc(30,50){10}{0}
  \Vertex(90,80){1.2}
  \Vertex(90,20){1.2}
  \ArrowLine(90,80)(120, 95)
  \ArrowLine(120,65)(105,72.5)
  \ArrowLine(105,72.5)(90,80)
  \Vertex(105,72.5){1.2}
  \ArrowLine(120, 5)(90,20)
  \ArrowLine(90,20)(120,35)
  \Photon(60,35)(105,72.5){2}{5}
  \put(87,46){$\ga$}
  \put(63,12){\PW}
  \put(40,24){\PW}
  \put(55,73){\PW}
  \end{picture}
  \quad
  \begin{picture}(160,100)(0,0)
  \ArrowLine(30,50)( 5, 95)
  \ArrowLine( 5, 5)(30, 50)
  \Photon(30,50)(90,80){-2}{6}
  \Photon(30,50)(90,20){2}{6}
  \Vertex(60,65){1.2}
  \GCirc(30,50){10}{0}
  \Vertex(90,80){1.2}
  \Vertex(90,20){1.2}
  \ArrowLine(90,80)(120, 95)
  \ArrowLine(120,65)(105,72.5)
  \ArrowLine(105,72.5)(90,80)
  \Vertex(105,27.5){1.2}
  \ArrowLine(120, 5)(90,20)
  \ArrowLine(105,27.5)(120,35)
  \ArrowLine(90,20)(105,27.5)
  \Photon(60,65)(105,27.5){-2}{5}
  \put(84,55){$\ga$}
  \put(63,78){\PW}
  \put(40,68){\PW}
  \put(55,16){\PW}
  \end{picture}
%
  \\[2ex]
  \unitlength .7pt\small\SetScale{0.7}
  \begin{picture}(240,100)(0,0)
  \ArrowLine(30,50)( 5, 95)
  \ArrowLine( 5, 5)(30, 50)
  \Photon(30,50)(90,80){2}{6}
  \Photon(30,50)(90,20){2}{6}
  \GCirc(30,50){10}{0}
  \Vertex(90,80){1.2}
  \Vertex(90,20){1.2}
  \ArrowLine(90,80)(120, 95)
  \ArrowLine(120,65)(105,72.5)
  \ArrowLine(105,72.5)(90,80)
  \ArrowLine(120, 5)( 90,20)
  \ArrowLine( 90,20)(120,35)
  \Vertex(105,72.5){1.2}
  \PhotonArc(120,65)(15,150,270){2}{3}
  \put(55,73){\PW}
  \put(55,16){\PW}
  \put(99,47){$\ga$}
  \DashLine(120,0)(120,100){6}
  \PhotonArc(120,35)(15,-30,90){2}{3}
  \Vertex(135,27.5){1.2}
  \ArrowLine(150,80)(120,95)
  \ArrowLine(120,65)(150,80)
  \ArrowLine(120, 5)(150,20)
  \ArrowLine(150,20)(135,27.5)
  \ArrowLine(135,27.5)(120,35)
  \Vertex(150,80){1.2}
  \Vertex(150,20){1.2}
  \Photon(210,50)(150,80){2}{6}
  \Photon(210,50)(150,20){2}{6}
  \ArrowLine(210,50)(235,95)
  \ArrowLine(235, 5)(210,50)
  \GCirc(210,50){10}{0}
  \put(177,73){\PW}
  \put(177,16){\PW}
  \end{picture}
  \ece
  \caption[]{Examples for virtual (top) and real (bottom) non-factorizable 
             corrections to \PW-pair production.}
  \label{fi:nf}
\efi%
In the DPA the NFC arise exclusively from the exchange or emission of photons 
with $E_\ga \lsim {\cal O}(\GV)$. Hard photons as well as other massive 
particles do not lead to double-resonant contributions. 

In the case of photon emission from a W boson
(see \reffi{FSRfig}), 
the split-up between 
factorizable and non-factorizable corrections can be achieved with the help 
of a partial-fraction decomposition of the two W-boson propagators 
separated by the photon~\cite{Be85}: 
\beq
\label{decomp}
  \frac{1}{[p^2-M^2][(p-k)^2-M^2]} 
  =
  \frac{1}{2(p\cdot k)}\,\Biggl( \frac{1}{(p-k)^2-M^2}
                               - \frac{1}{p^2-M^2} \Biggr),
\eeq
where $M^2=\MW^2-\ri\MW\GW$, $k$ is the momentum of the emitted photon,
and $p$ is the momentum of the \PW~boson before emission. In this way one 
obtains a sum of two resonant W-boson propagators multiplied by an ordinary 
on-shell eikonal factor. This decomposition allows a gauge-invariant split-up 
of the real-photon matrix element in terms of one contribution where the 
photon is effectively emitted from the production part, and another two where 
the photon is effectively emitted from one of the two decay parts. 
The squares of the three contributions can be identified as factorizable 
corrections, whereas the interference terms are of non-factorizable nature. 
The same type of split-up can be performed for the corresponding virtual 
corrections.

The factorizable corrections 
have the nice feature that they can be expressed 
in terms of corrections to on-shell subprocesses, \ie the production of two
on-shell gauge bosons ($\PV_1\PV_2$) and their subsequent on-shell
decays. In this way the well-known on-shell radiative corrections to
the production and decay of pairs of gauge bosons (see \citere{WWreview}
and references therein) appear as basic building blocks of the
factorizable corrections. For instance, for the virtual factorizable
corrections in DPA one finds
\beq\label{mvirt}
  \M_{\mathrm{f}} = \sum_{V\; \mathrm{pol.}}
      \frac{\M^{\Pep\Pem\to \PV_1 \PV_2} \M^{\PV_1\to f_1\bar f_2} 
      \M^{\PV_2\to f_3\bar f_4}}{(q_1^2-M^2)(q_2^2-M^2)},
\eeq
with $M^2=\MV^2-\ri\MV\GV$ and $q_{1,2}^2$ the invariant masses squared of the
bosons $V_{1,2}$. The off-shell character of the reaction is reflected by the 
occurrence of the two Breit--Wigner resonances. 

\subsection{Factorizable real-photon corrections in double-pole approximation}

As indicated above, the factorizable real-photon corrections are characterized 
by their close relation to on-shell subprocesses. In this context three 
regimes for the photon energy play a role:
\begin{itemize}
  \item for hard photons [$E_{\gamma}\gg \GV$] the Breit--Wigner poles of the 
        gauge-boson resonances before and after photon radiation (see $s'_V$ 
        and $s_V$ in \reffi{FSRfig}) are well separated in phase space. 
        This leads to three {\it distinct} regions of on-shell contributions, 
        where the photon can be assigned unambiguously to the 
        gauge-boson-pair-production 
        subprocess or one of the two decays. 
        Therefore, the double-pole residue can
        be expressed as the sum of the three on-shell 
        contributions without increasing the intrinsic error of the DPA. 
        Note that in the same way it is also possible to experimentally assign 
        the photon to one of the subprocesses, since misassignment errors are 
        suppressed.
  \item for ``semi-soft'' photons [$E_{\gamma} = {\cal O}(\GV)$] the 
        Breit--Wigner poles are relatively close together in phase space, 
        resulting in a  substantial overlap of the line shapes. The assignment 
        of the photon is now subject to larger errors, and the DPA has to be 
        applied with caution. 
  \item for soft photons [$E_{\gamma}\ll \GV$] the Breit--Wigner poles are on 
        top of each other, resulting in a pole-scheme expansion that is 
        identical to the one without the photon.
\end{itemize}

If the photon is integrated out, the reduction of the five-particle phase space
to a four-particle one introduces a parametrization dependence. 
As we will see in the following, the size of the radiative corrections depends 
strongly on the choice of parametrization of the four-particle phase space. 

When one produces two resonances, or one resonance and a stable particle, the 
line shape of such a resonance is measured from the invariant-mass 
distribution of its decay products. This has to be contrasted with the Z 
resonance at LEP1, which is defined as a function of the centre-of-mass energy
squared. Depending on how one measures the invariant-mass distribution, 
different sources of Breit--Wigner distortions can be identified. At LEP1 such 
a distortion is caused by initial-state radiation (ISR)~\cite{zreport}, but in
our case also final-state radiation (FSR) can be responsible.
 
In \citere{fact-fsr} it has been shown that such a FSR-induced distortion is a 
general property of resonance-pair reactions, irrespective of the adopted 
scheme for implementing the finite-width effects. The only decisive factor for 
the distortion to take place is whether the virtuality of the unstable 
particle is defined with or without the radiated photon (see \reffi{FSRfig}). 
\bfi
  \bce
  \unitlength .7pt\small\SetScale{0.7}
  \begin{picture}(150,90)(0,15)
    \ArrowLine(43,43)(25,25)          \Text(8,25)[lc]{\Pep}
    \ArrowLine(25,75)(43,57)          \Text(8,77)[lc]{\Pem}
    \Photon(50,50)(110,80){1}{9}
    \LongArrow(61,63)(71,68)          \Text(62,74)[]{$p$}
                                      \Text(76,52)[]{\PV} 
    \LongArrow(85,75)(95,80)          \Text(88,89)[]{$p\!-\!k$} 
                                      \Text(102,66)[]{\PV}    
    \Line(50,50)(75,40)
    \Line(50,50)(67,30)
    \ArrowLine(110,80)(135,105)
    \ArrowLine(135,55)(110,80)
    \Vertex(110,80){1.5}
    \Photon(80,65)(105,40){-1}{6}      \Text(105,30)[]{$\ga$}
    \Vertex(80,65){1.5}
    \LongArrow(104,51.5)(111,44.5)    \Text(117,53.5)[]{$k$}
    \GCirc(50,50){10}{0}
  \end{picture}
  \ece 
  \caption[]{Photon radiation from an unstable particle V. Virtualities:
             $s_{\PV} = (p-k)^2$ and $s'_{\PV} = p^2$.}
  \label{FSRfig}
\efi%
Upon integration over the photon momentum, the former definition ($s'_{\PV}$) 
is free of large FSR effects from the \PV-decay system. It can only receive 
large corrections from the other (production or decay) stages of the process. 
The latter definition ($s_{\PV}$), however, does give rise to large FSR 
effects from the \PV-decay system. In contrast to the LEP1 case, where the 
ISR-corrected line shape receives contributions from effectively {\it lower} 
Z-boson virtualities, the $s_{\PV}$ line shape receives contributions from 
effectively {\it higher} virtualities $s'_{\PV}$ of the unstable particle. 
As was argued above, only sufficiently hard photons ($E_{\ga}\gg \GV$) can be 
properly assigned to one of the on-shell production or decay stages of the 
process in the DPA. For semi-soft photons [$E_{\ga} = {\cal O}(\GV)$], however,
the assignment lacks a solid motivation and an invariant-mass definition in 
terms of the decay products without the photon seems more natural. As was 
pointed out in \citere{fact-fsr}, exactly these semi-soft photons are 
responsible for the FSR-induced distortion effects. The hard FSR photons move 
the virtuality $s'_{\PV}$ of the unstable particle far off resonance for 
near-resonance $s_{\PV}$ values, resulting in a suppressed contribution to the 
$s_{\PV}$ line shape. This picture fits in nicely with the negligible overlap 
of the three on-shell double-pole contributions for hard photons, discussed 
above.

The size of the FSR-induced distortion can be parametrized in terms of the
shift in the peak position and the corresponding peak reduction factor with
respect to the lowest-order line shape. These parameters depend strongly on the
type of unstable particle and the precise experimental definition of the 
associated invariant mass.\footnote{In realistic event-selection procedures a
   minimum opening angle between the decay products and the photon is required 
   for a proper identification of all particles. This affects the integration
   range of the photon momentum and therefore the size of the distortion.} 
The shift in the peak position generally amounts to several times $-10\MeV$.
As extreme values we mention $-110\MeV$ for an unstable Z boson that decays
as $\PZ\to \Pep\Pem$ and $-45\MeV$ for an unstable W boson that decays as 
$\PW\to \Pe\Pne$, assuming no minimum opening angle between the leptons and 
the photon. The peak reduction factor lies in the range 0.95\,--\,0.75.
All this stresses the importance of a proper inclusion of FSR effects in the
experimental method for extracting the gauge-boson masses from the 
reconstructed line shapes.

As an example we display in \reffi{distortion} the large FSR-induced 
distortion effects for the double-resonance toy process 
$\Pnmu\Panmu \to \PZ\PZ \to \Pem\Pep\Pnta\Panta$.
\bfi
  \unitlength 1cm
  \bce
  \begin{picture}(8,5)
  \put(0,3.5){\makebox[0pt][c]
             {\boldmath $\frac{\rd\sigma}{\rd M_1^2\rd M_2^2}$}} 
  \put(0,2.5){\makebox[0pt][c]
             {\boldmath $\bigl[\mbox{\bf \scriptsize pb}/
                               \mbox{\bf \scriptsize GeV}^{4}\bigr]$}}
  \put(8,0.2){\makebox[0pt][c]{\boldmath $\scriptstyle{M_1}\,
                               [\mbox{\bf \scriptsize GeV}]$}}
  \put(0.5,-3){\includegraphics{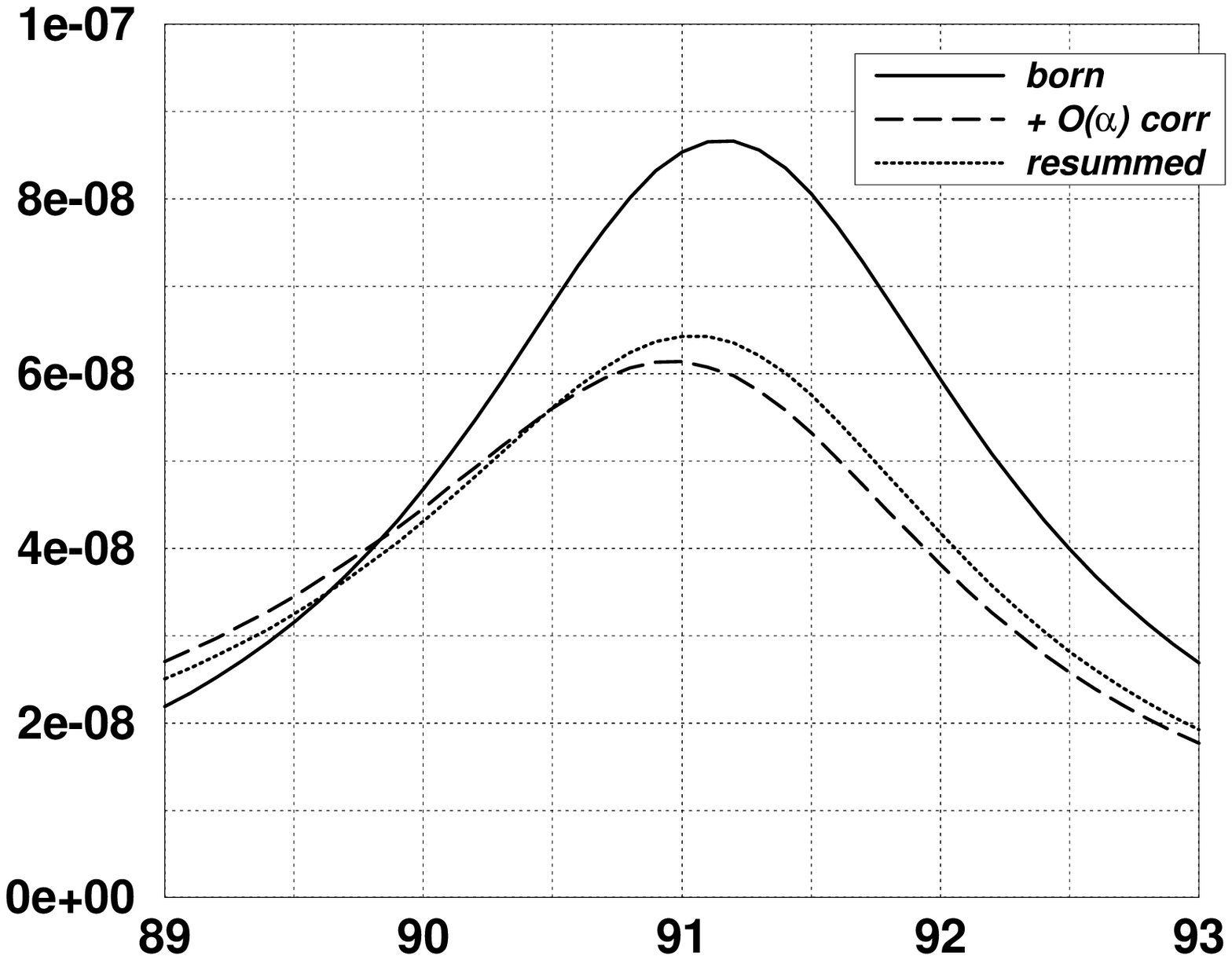}}
  \end{picture}
  \ece
  \caption[]{The FSR-induced distortion of the line shape
             $\protect\rd\sigma/(\protect\rd M_1^2\,\protect\rd M_2^2)$ 
             corresponding to the toy process 
             $\Pnmu\Panmu \to \PZ\PZ \to \Pem\Pep\Pnta\Panta$ 
             for $M_2=\MZ$. Here $M_1$ and $M_2$ stand for the invariant 
             masses of the $\Pem\Pep$ and $\Pnta\Panta$ pair, respectively.
             Centre-of-mass energy: $\protect\sqrt{s}=200\GeV$. No minimum
             opening angle between the photon and $\Pe^\pm$. 
             Plot taken from \citere{fact-fsr}.}
\label{distortion}
\efi%
Apart from being exactly calculable, this process is free of the usual 
gauge-invari\-ance problems and it only receives QED corrections from the  
$\PZ \to \Pep\Pem$ decay. As such it is well suited for showing the salient
features of FSR. Like at LEP1, the size of the ${\cal O}(\alpha)$ corrections 
indicates the need for a resummation of soft-photon effects. A justification 
of this can be found in the fact that only semi-soft photons lead to the 
distortion. The resummation factor $\delta_{\mbox{\scriptsize res}}$ that 
multiplies the lowest-order double-invariant-mass distribution in 
\reffi{distortion} can be accurately approximated by   
\beqar
\label{FSRresummed}
  \delta_{\mbox{\scriptsize res}} &=& \bigl(1+\frac{3}{4}\,\be)
           \int\limits_{1}^{\infty} \rd\zeta\,\be\,(\zeta-1)^{\be-1}\,
           \frac{|D_{\PZ}(M_1^2)|^2}{|D_{\PZ}(\zeta M_1^2)|^2}
           \nonumber \\[1mm]  
                                  &=& \bigl(1+\frac{3}{4}\,\be)
           \,\frac{\pi\be}{\sin(\pi\be)}\,
           \real\Biggl[\frac{\ri D_{\PZ}^*(M_1^2)}{\MZ\Ga_{\PZ}}\,
           \biggl(\frac{D_{\PZ}(M_1^2)}{\MZ^2}\biggr)^{\be}\,\Biggr], 
\eeqar
with $\,\be = (2\alpha/\pi)\,[\log(\MZ^2/\Me^2)-1]\,$ and 
$\,D_{\PZ}(p^2) = p^2-\MZ^2+\ri\MZ\Ga_{\PZ}$.\footnote{Equation 
    (\ref{FSRresummed}) equally applies to other unstable particles and decay
    products, provided the definitions of the resonance parameters 
    ($\MZ,\Ga_{\PZ}$) and $\be$ are properly adjusted. For instance,
    a $\PW\to \mu\Pnmu$ decay requires $\MZ \rightarrow \MW$,
    $\Ga_{\PZ} \rightarrow \GW$, $\Me \rightarrow \Mmy$ and 
    $\alpha \rightarrow \alpha/2$.}
The fact that the FSR-corrected 
line shape receives contributions from effectively {\it higher} virtualities is
reflected by the parameter $\zeta$, which takes on values above unity. Apart
from the inversion of $\zeta$, the situation is the same as for ISR at LEP1.
This analogy is also confirmed by simple rules of thumb for the peak position 
and peak reduction factor (\cf \citeres{fact-fsr,thumb}).

\subsection{Non-factorizable corrections}

The NFC to gauge-boson pair production have been considered by several authors
in recent years. First it was shown that the NFC vanish in inclusive 
quantities~\cite{Fa94,Me94}, \ie if the invariant masses of both gauge bosons 
are integrated out. The NFC to differential distributions in \PW-pair
production were first calculated in \citere{Me96}, but the analytical results 
were given only in an implicit form and the numerical evaluation was 
restricted to a special phase-space configuration. Recently two groups have 
independently provided both the complete formulae and an adequate numerical 
evaluation for the leptonic final state~\cite{Be97,De98}. While these latter 
two calculations agree analytically and numerically, they deviate from the 
results of \citere{Me96}. Subsequently, also numerical results for other final 
states in \PW-pair production and \PZ-pair production were 
investigated~\cite{De97}. In the following, we discuss the results based on 
\citeres{Be97,De98,De97}.

The manifestly non-factorizable diagrams (see \reffi{fi:nf} for examples) 
are not gauge-invariant. A gauge-invariant definition of the NFC can be given 
in different ways. One possibility, used in \citere{Be97}, is based on the 
fact that only soft and semi-soft photons contribute in the DPA. The associated
matrix element can be written as a product of the lowest-order matrix element 
times conserved soft-photon currents, which can be decomposed into one
production and two decay currents with the help of \refeq{decomp}.
The NFC are then defined as all interferences between the three currents.

A second definition~\cite{De98} makes use of the gauge independence of the
complete double-resonant corrections. The factorizable double-resonant 
corrections are given by the product of gauge-invariant on-shell matrix 
elements for gauge-boson pair production and gauge-boson decays, and the
(transverse parts of the) gauge-boson propagators [see \refeq{mvirt}]. 
Therefore, the double-resonant NFC can be defined by subtracting the 
double-resonant factorizable corrections from the complete double-resonant 
corrections. It turns out that both definitions are equivalent in the DPA and, 
for charged bosons, include parts of the diagram shown in \reffi{mixeddiagram}
(\cf \citere{Fa94}).  
\bfi 
  \bce {\unitlength  .7pt\small\SetScale{0.7}
  \begin{picture}(170,100)(0,0)
  \ArrowLine(30,50)( 5, 95)
  \ArrowLine( 5, 5)(30, 50)
  \Photon(30,50)(90,80){2}{6}
  \Photon(30,50)(90,20){2}{6}
  \Photon(70,30)(70,70){2}{4}
  \Vertex(70,30){1.2}
  \Vertex(70,70){1.2}
  \GCirc(30,50){10}{0}
  \Vertex(90,80){1.2}
  \Vertex(90,20){1.2}
  \ArrowLine(90,80)(120, 95)
  \ArrowLine(120,65)(90,80)
  \ArrowLine(120, 5)( 90,20)
  \ArrowLine( 90,20)(120,35)
  \put(76,47){$\ga$}
  \put(45,68){\PW}
  \put(45,22){\PW}
  \put(72,83){\PW}
  \put(71,8){\PW}
  \end{picture} }
  \ece
  \caption[]{Diagram contributing both to factorizable and non-factorizable 
             corrections.}
  \label{mixeddiagram}
\efi 
 
Since only soft and semi-soft photons are relevant, each virtual diagram 
contributing to the NFC reduces to the corresponding lowest-order matrix 
element times a correction factor that involves besides kinematical variables 
only a scalar integral. After evaluation of this scalar integral, an expansion 
of the residue around the poles at $\kV^2 = \MV^2$ can be performed. 
The constant finite width can be introduced either before integration or in 
the final result. It regularizes besides the resonant propagators also 
logarithms of the form $\log(\kV^2-\MV^2)$.
 
The real-photon NFC can be calculated in a similar way. The integration region 
for the energy of the real photon can be extended to infinity without changing 
the result. However, when integrating over the photon momentum, the complete 
parametrization of phase space has to be specified~\cite{De98}. In analogy to 
the factorizable corrections, also the real-photon NFC are not universal
but depend on the choice of this parametrization, even in DPA. We follow the 
usual procedure and take the invariant masses of decay fermion pairs as 
independent variables.
 
After combining real and virtual contributions, the NFC yield a simple 
polarization-independent correction factor to the lowest-order cross-section:
\beq
\rd\si_\nf = \de_\nf \, \rd\si_\born.
\eeq
For the processes
\beq
  \label{process}
  \Pep(p_+) + \Pem(p_-) \;\to\; \PV_1(q_1) + \PV_2(q_2)
  \;\to\; f_1(k_1) + \bar f_2(k_2) + f_3(k_3) + \bar f_4(k_4),
\eeq
with $\PV_1\PV_2 = \PW\PW$ or $\PZ\PZ$, this factor can be written 
as~\cite{De98}
\beq\label{nfcorrfac}
  \delta_\nf(k_1,k_2;k_3,k_4) = \sum_{a=1,2 \atop b=3,4} \, (-1)^{a+b+1} 
                                \, Q_a Q_b \,\frac{\alpha}{\pi} \, 
                                \real\{\Delta(q_1,k_a;q_2,k_b)\}.
\eeq
Here $Q_i$ $(i=1,2,3,4)$ denotes the relative charge of fermion $f_i$ and 
$q_1=k_1+k_2$, $q_2=k_3+k_4$ the momenta of the virtual bosons. Each 
term in the sum of \refeq{nfcorrfac} corresponds to the photon exchange
between two specific final-state fermions that originate from different bosons.
The function $\De$ is explicitly given in \citere{De98} and, in a different 
notation, in \citere{Be97}. It has in DPA the important property:
($K_i \equiv q_i^2-\MV^2+\ri\MV\GV$)
\beq\label{relation}
  \De + \De\bigg|_{K_1 \,\to\, -\,K_1^*} + \De\bigg|_{K_2 \,\to\, -\,K_2^*}
  + \De\bigg|_{K_{1,2} \,\to\, -\,K_{1,2}^*} = 0. 
\eeq

The final result for the NFC has the following general features:
\begin{itemize}
\item Photon-exchange contributions between initial and final states cancel 
      between virtual and real corrections, leaving behind the cross-talk 
      between the two decay systems as only contributions. Consequently, the 
      NFC are independent of the production angle of the gauge bosons and 
      \refeq{nfcorrfac} is applicable to other initial states like 
      $q\bar{q}$ or $\ga\ga$. 
\item The mass-singular logarithms that are present in individual contributions
      cancel, and the complete NFC are free of mass singularities. 
      As a consequence, the typical order of magnitude of the NFC is 
      $\al/\pi\sim 0.3\%$.
\item Equation \refeq{relation} implies that the NFC vanish if both virtual 
      gauge bosons are on-shell. This leads to a suppression with 
      respect to the factorizable corrections. For single-invariant-mass 
      distributions the suppression factor is $(q_{1,2}^2-\MV^2)/(\MV\GV)$.
      This is of order one in the vicinity of the resonance, \ie for 
      $q_{1,2}^2-\MV^2 = {\cal O}(\MV\GV)$, ensuring that the NFC are
      genuine double-pole contributions.
\item As can be seen from \refeq{relation}, the NFC vanish if both invariant 
      masses are integrated over. Thus, they vanish for pure angular 
      distributions and therefore have no impact on standard TGC studies. 
\item Finally, the NFC vanish at high energies. This can be easily illustrated
      by the result for the production of one stable and one unstable charged 
      particle with equal masses, where the correction factor 
      becomes~\cite{Me96}
      $$
      \de_{\nf} = -\,\al\,\frac{(1-\be)^2}{\be}
                  \arctan\left(\frac{q^2-M_V^2}{\MV\GV}\right)
      \quad \mbox{with} \quad \be =\sqrt{1-4\MV^2/s}.
      $$
      This simple formula describes qualitatively also the realistic case of 
      the single-invariant-mass distribution in the case of two unstable 
      particles.
\end{itemize}
These features indicate that the NFC are less important than the factorizable
corrections. But their actual relevance depends on the experimental accuracy 
and the observable under consideration.

We first consider the NFC to $\Pep\Pem \to \PW\PW \to 4f$. 
In \reffi{energyplot} we show the NFC to the single-invariant-mass 
distribution $\rd\sigma/\rd M_1$ of the process 
$\Pep\Pem \to \PW\PW \to \Pne\Pep\Pem\Pane$ for various centre-of-mass 
energies. With $M_{1,2}$ we denote the invariant masses of the first and second
fermion--antifermion pairs, respectively. The NFC are roughly 1\% for 
$\sqrt{s}=172\GeV$ and decrease fast with increasing energy.
\bfi
  \centerline{ 
  \setlength{\unitlength}{1cm}
  \begin{picture}(6,5.3)
  \put(0,0){\includegraphics{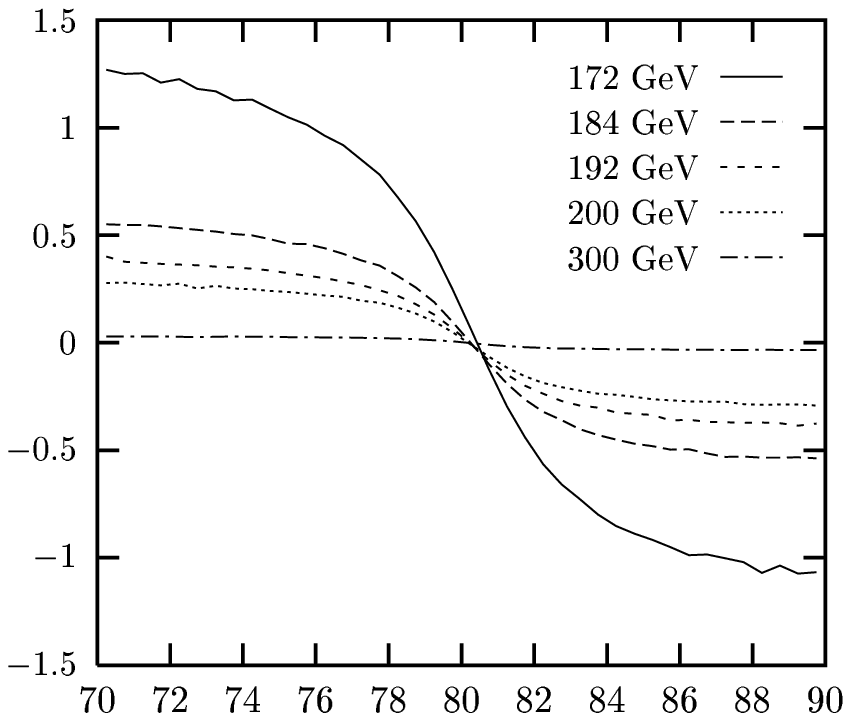}}
  \put(-0.3,3.7){\makebox(1,1)[c]{$\de_\nf$}}
  \put(-0.3,2.9){\makebox(1,1)[c]{\scriptsize [\%]}}
  \put(6.3,0.4){\makebox(1,1)[cc]{{$\scriptstyle{M_1}\,
                                   [\mbox{\scriptsize GeV}]$}}}
  \end{picture}
  }
  \caption[]{Relative NFC to the single-invariant-mass distribution
             $\protect\rd\sigma/\protect\rd M_1$ for
             $\Pep\Pem \to \PW\PW \to \Pne\Pep\Pem\Pane$.
             Plot taken from \citere{De98}.}
\label{energyplot}
\efi%
They distort the invariant-mass distribution and thus in principle influence 
the determination of the \PW-boson mass from the direct reconstruction of the 
decay products. The corresponding mass shift can be estimated by the
displacement of the maximum of the single-invariant-mass distribution caused 
by the corrections shown in \reffi{energyplot}. This displacement turns out to 
be of the order of 1\,--\,2\,MeV, \ie small compared to the LEP2 accuracy.

In \reffi{Winvmass} we compare the results for the single-invariant-mass 
distribution $\rd\sigma/\rd M_1$ for various final states, once calculated 
using the code of \citere{De98} and once using the code of \citere{Be97}.
\bfi
  \centerline{
  \setlength{\unitlength}{1cm}
  \begin{picture}(11.5,5.5)
  \put(0,0){\includegraphics{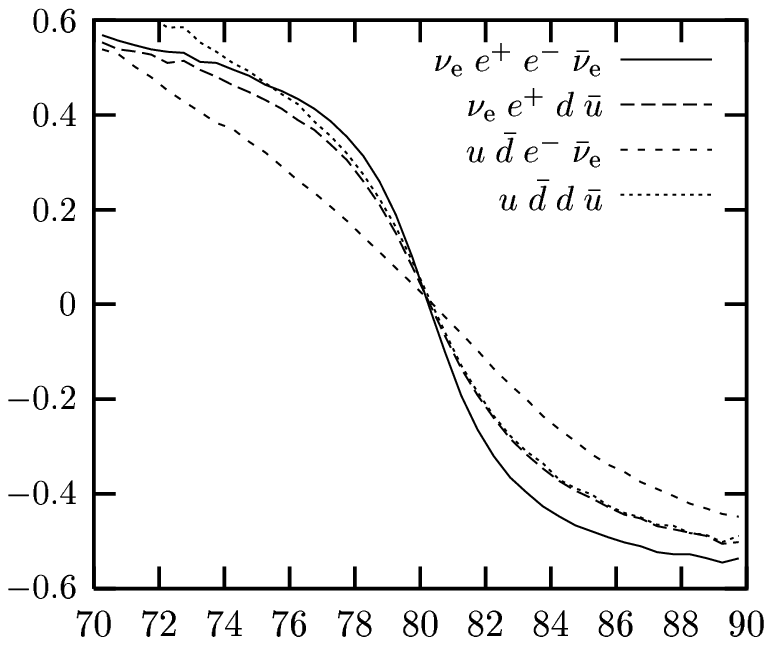}}
  \put(0,0){\includegraphics{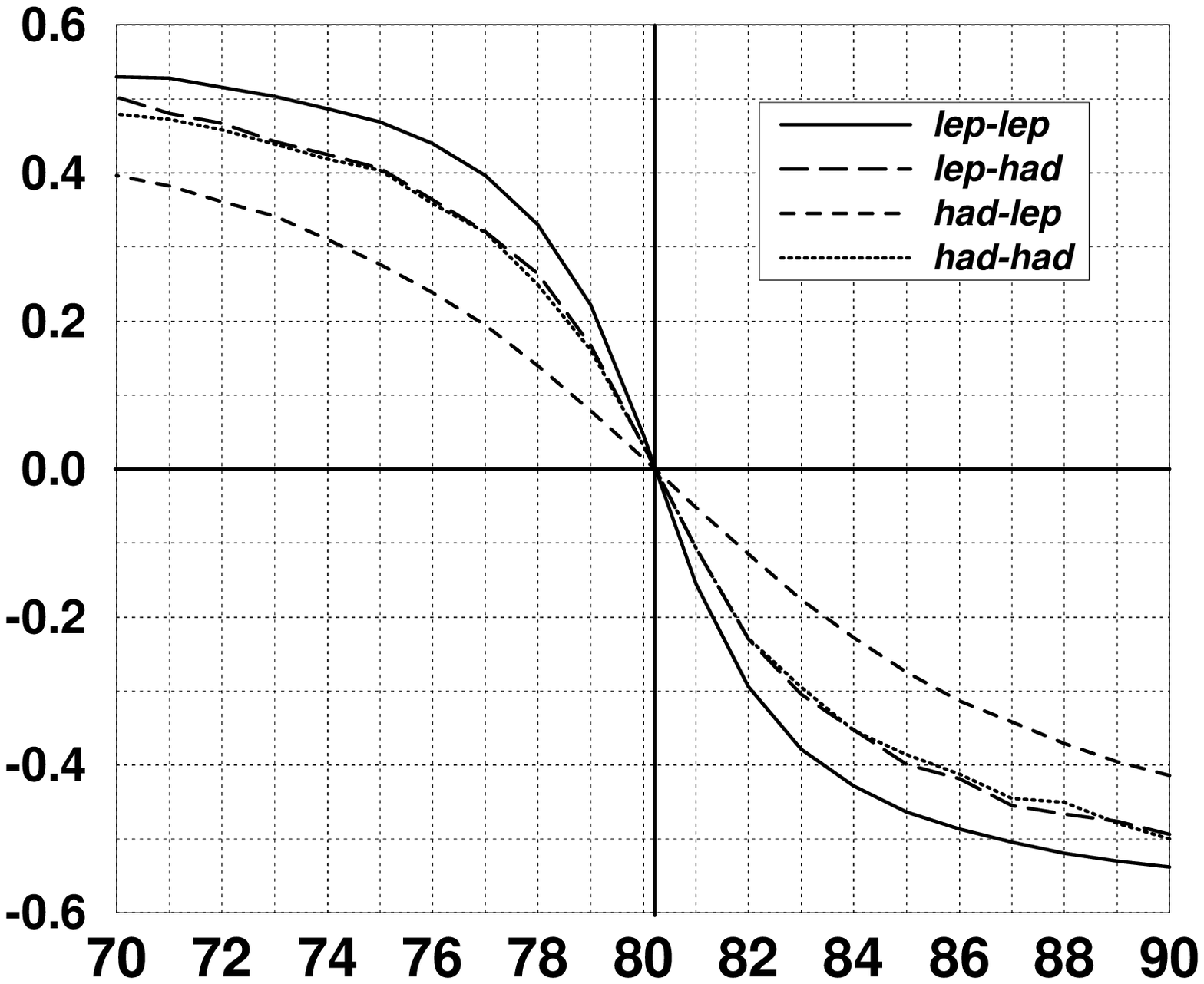}}
  \put(-0.5,3.9){\makebox(1,1)[c]{$\de_\nf$}}
  \put(-0.5,3.1){\makebox(1,1)[c]{\scriptsize [\%]}}
  \put(2.9,-0.1){\makebox(1,1)[cc]{{$\scriptstyle{M_{1}}\,
                                    [\mbox{\scriptsize GeV}]$}}}
  \put(8.5,-0.1){\makebox(1,1)[cc]{{\boldmath $\scriptstyle{M_1}\,
                                    [\mbox{\bf \scriptsize GeV}]$}}}
  \end{picture}
  }
  \caption[]{Relative NFC to the single-invariant-mass distribution 
             $\protect\rd\sigma/\protect\rd M_1$ for 
             $\Pep\Pem \to \PW\PW \to 4f$ with different final states.
             Left: results from \citere{De97}. Right: results based on 
             \citere{Be97}. Centre-of-mass energy: $\protect\sqrt{s}=184\GeV$.}
  \label{Winvmass}
\efi%
The NFC are similar for all final states. The NFC to the distribution 
$\rd\sigma/\rd M_2$ can be derived from the results for $\rd\sigma/\rd M_1$ by
a CP transformation of the final state~\cite{De97}. The results for the
single-invariant-mass distributions agree very well between the two different 
calculations, in particular for large invariant masses and for invariant 
masses close to the \PW-boson mass, which dominate the cross-sections. 
The discrepancies for smaller invariant masses are of the order of the 
non-double-resonant corrections and are due to different 
implementations of the corrections. While in the numerical 
evaluations of \citere{Be97} the phase space and the Born matrix element are 
taken entirely on-shell, these are taken off-shell in the evaluation of 
\citeres{De98,De97}. 

Results for the dependence of the NFC on angles and energies of
the final-state fermions can be found in \citeres{De98,De97}. These NFC
are typically of the order of one percent, \ie somewhat larger than
for the invariant-mass distributions.
The largest effects of some per cent can be observed near the edges of
phase space, where statistics is limited.  

Next we consider the NFC to $\Pep\Pem \to \PZ\PZ \to 4f$. 
\bfi
  \centerline{
  \setlength{\unitlength}{1cm}
  \begin{picture}(11.5,5.5)
  \put(0,0){\includegraphics{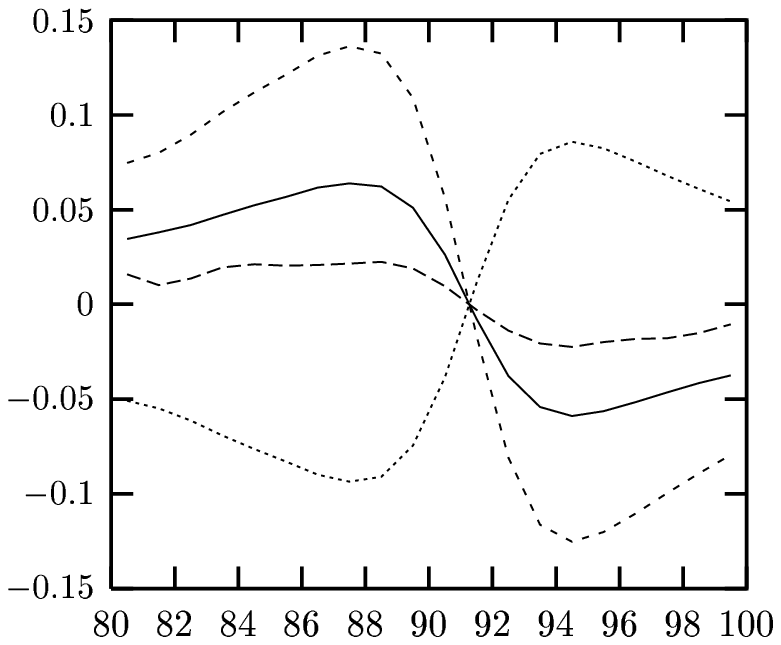}}
  \put(0,0){\includegraphics{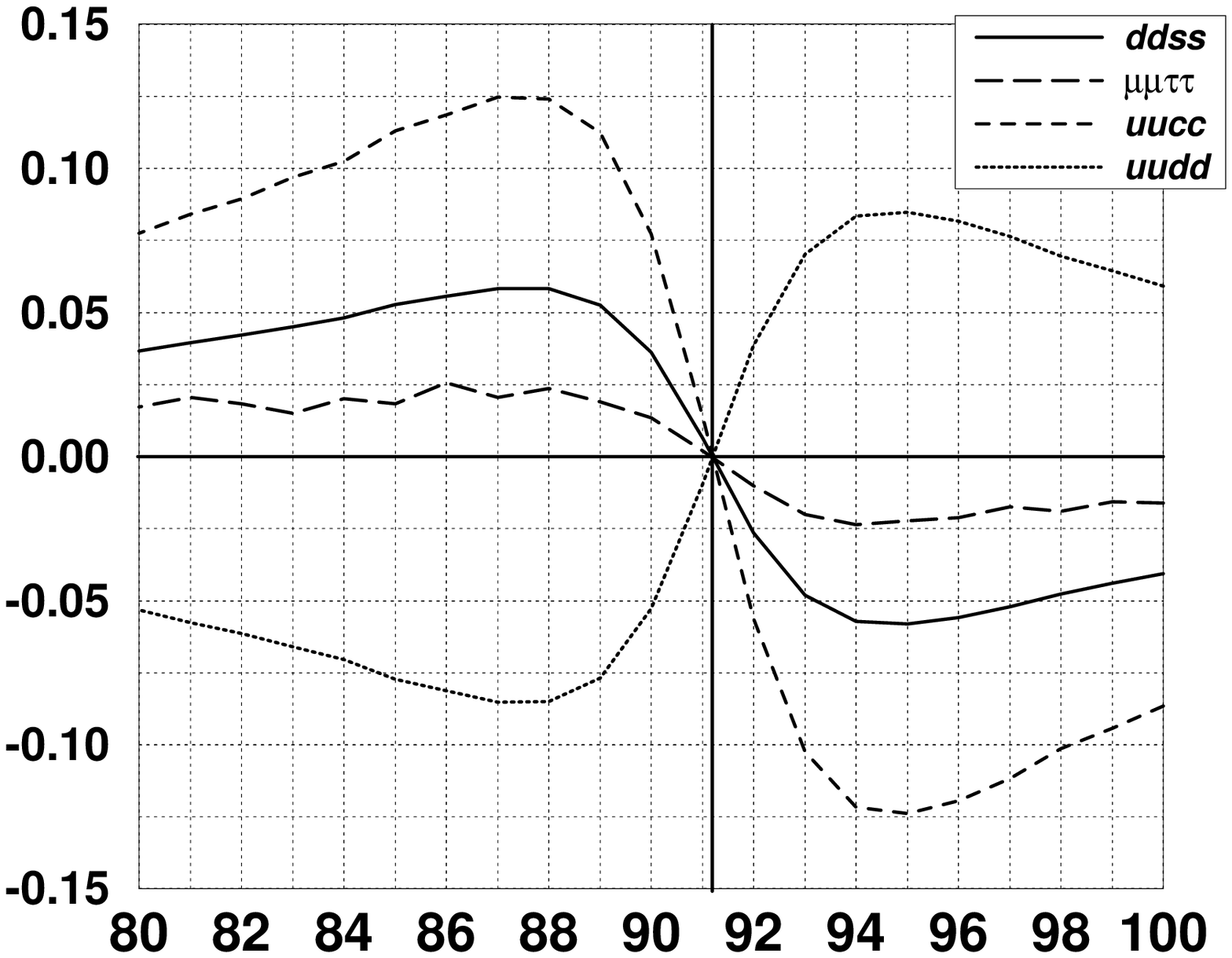}}
  \put(-0.5,3.9){\makebox(1,1)[c]{$\de_\nf$}}
  \put(-0.5,3.1){\makebox(1,1)[c]{\scriptsize [\%]}}
  \put(2.9,-0.1){\makebox(1,1)[cc]{{$\scriptstyle{M_{1,2}}\,
                                    [\mbox{\scriptsize GeV}]$}}}
  \put(8.5,-0.1){\makebox(1,1)[cc]{{\boldmath $\scriptstyle{M_{1,2}}\,
                                    [\mbox{\bf \scriptsize GeV}]$}}}
  \end{picture}
  }
  \caption[]{Relative NFC to the single-invariant-mass distributions 
             $\protect\rd\sigma/\protect\rd M_{1,2}$ for 
             $\Pep\Pem \to \PZ\PZ \to 4f$ with different final states.
             Left: results from \citere{De97}. Right: results based on 
             \citere{Be97}. Centre-of-mass energy: $\protect\sqrt{s}=192\GeV$.}
  \label{Zinvmass}
\efi%
Figure~\ref{Zinvmass} shows the NFC to the single-invariant-mass distribution 
$\rd\sigma/\rd M_1$, which is identical to $\rd\sigma/\rd M_2$. Again the 
results of \citere{De97} agree well with those based on
\citere{Be97}. For opposite 
signs of $Q_1$ and $Q_3$ the sign of the corrections is reversed with respect 
to $\Pep\Pem \to \PW\PW \to 4f$. The corrections by themselves are 
very small and phenomenologically unimportant. The smallness of these 
corrections results from the fact that $\de_{\mathrm{nf}}$ is symmetric in 
$k_1\leftrightarrow k_2$ and $k_3\leftrightarrow k_4$ after
integration over the decay angles. For all observables that involve an
integration over the phase space that respects this symmetry, the NFC are 
suppressed either by the charges or the vector couplings of the final-state 
fermions~\cite{De97}.

If the decay angles are not integrated out, this suppression does not
apply.  In fact, owing to the presence of four charged final-state
fermions, the NFC to angular \PZ-pair distributions are enhanced with
respect to the \PW-pair case by a factor of roughly four for purely
leptonic final states, and can amount to up to 10\%~\cite{De97}. Note,
however, that the cross-section for \PZ-pair production is only one
tenth of the \PW-pair production cross-section.

\section{Summary}

For a gauge-invariant treatment of finite-width effects in lowest-order 
off-shell gauge-boson pair production two consistent schemes exist, \ie the
fermion-loop scheme and the pole scheme. For the calculation of the radiative
corrections the latter scheme offers a promising framework in the form of the 
double-pole approximation. This approximation results in a split-up of the
corrections in so-called factorizable and non-factorizable corrections.
As far as the factorizable corrections are concerned, in particular the
photonic ones are crucial for coming up with adequate theoretical predictions. 
While in this context the importance of initial-state radiation is evident and 
commonly acknowledged, also the Breit--Wigner distortions induced by 
final-state radiation should be taken into account properly. 
The non-factorizable corrections to \PW-pair and \PZ-pair production are small 
with respect to the experimental accuracy at LEP2. As a consequence, the
factorizable corrections are sufficient for theoretical predictions for LEP2. 
The size of the non-factorizable corrections might, however, compete with the
expected experimental accuracy at future linear $\Pep\Pem$ colliders
with higher luminosity.

\section*{Acknowledgement}
We thank A.P.~Chapovsky, F.A.~Berends, S.~Dittmaier and M.~Roth for helpful
discussions and for stimulating collaborations.
W.B. would also like to acknowledge the support by a fellowship of the Royal 
Dutch Academy of Arts and Sciences.

\end{document}